**Gravitational energy released when Earth quakes: Heat supply for island-arc volcanism**

Benjamin F. Chao,[1*] Sheng-An Shih,[2] Frédéric Deschamps,[2] Eh Tan[2]

[1]College of Geodesy and Geomatics, Shandong University of Science and Technology, Qingdao, China

[2]Institute of Earth Sciences, Academia Sinica, Taipei, Taiwan

*Corresponding author email: bfchao@earth.sinica.edu.tw

**Abstract**

We address a long-standing paradox in plate tectonics concerning the thermodynamics of its very origin: What supplies the thermal energy for the intense island-arc volcanism at the subduction zones, the supposedly cool end of the mantle convection? We surmise a first-principle physics model tracking the gravitational energy ($E_g$)-heat conversion within the mantle convection. The scenario points to a great amount of $E_g$ *release* of ~10 TW induced by the thrust-faulting earthquakes that occur predominantly in the upper mantle at the subduction zones. Generally ~1000 times greater than the seismic energy release or nearly a quarter of the total terrestrial heat flow, this $E_g$ release, upon conversion into heat in the irreversible convection process, supplies amply for the island-arc volcanism. Meanwhile similar but opposite and less vigorous $E_g$ *gains* at ~2 TW occur at the diverging plate boundaries via the normal-faulting earthquakes. This model resolves the said paradox, and more importantly provides a new paradigm pointing to an active role of seismicity in the plate tectonic energetics and evolution of the convecting mantle.

Key words: Subduction volcanism, mantle convection, earthquake-induced gravitational energy, heat source

## 1. Introduction

Planet Earth is a thermodynamic machine that undergoes irreversible evolutions with time, as has been for billions of years since born. The primary enduring consequence under the second law of thermodynamics is a net transport of the internal thermal energy or heat, originated from gravitational accretion plus radiogenic heating, outward to the surface and eventually lost to space by radiation.

Thanks to its physical configuration, the principal thermodynamic evolution of the present-day Earth consists mainly of the mantle convection manifesting as the plate tectonics (e.g., Turcotte and Shubert, 2002; Anderson, 2007). In a generic Rayleigh-Benard convection the gravitation plays the deciding role (the buoyancy is simply “negative gravity” when immersed in hydrostatic pressure originating ultimately from gravitation), whereby the single most fundamental energetic process within the convection is the cycling exchange of the gravitational potential energy, $E_g$, with the viscous heat dissipation. Yet in the case of the Earth's mantle convection, the attention paid to the energetics has remained scant in the geophysical literature (Verhoogen, 1980; Rubincam, 1979; Jones et al., 1996; Morgan et al., 2016; Chao et al., 2026).

In this paper we examine the plate tectonic energetics and surmise an $E_g$-heat conversion mechanism that bears significantly on the plate subduction processes in association with thrust earthquakes. The mechanism at the same time provides a ready resolution to the long-standing paradox concerning the energy source that supplies for the island-arc volcanism at the subduction zones, the supposedly cool end of the mantle convection cycle.

## 2. Earthquakes and Gravitational Energy

As a key component of plate tectonics, the surficial lithosphere is required naturally to respond to, or to accommodate, the grand thermodynamic evolution, either passively or actively. Being brittle, the lithosphere breaks when forced to deform, doing so at certain weak localities from time to time, and the Earth quakes. Seismology studies the earthquakes typically in terms of the elastic forces. In view of the *energetics*, however, earthquakes stand for the way decided of the lithosphere of present-day Earth to accommodate the Earth's ongoing thermodynamic evolution.

Focusing on energetics, an earthquake faulting turns the temporarily stored non-hydrostatic elastic strain energy into seismic wave energy which dissipates into heat eventually, plus comparable amounts of frictional and fracture heat generated locally upon faulting (Dahlen, 1977; Kanamori, 1977). Today the global seismic energy release by all earthquakes amounts to a mere average power of ~10 gigawatt as tallied from, say, the Global Centroid Moment Tensor (GCMT) Catalogue.

That accounts for all that would happen energy-wise with respect to an earthquake if the Earth were non-gravitating. In the gravitating Earth, however, an earthquake takes on a far more significant energetic role by tapping into the huge reservoir that is Earth's $E_g$. An earthquake reveals (per waveform inversion) its focal source mechanism (with magnitude) through the radiated seismic waves, based on which the earthquake-generated displacement field $\mathbf{u}(\mathbf{r})$ in the Earth can be calculated using, say, the normal-mode summation technique or elastic dislocation theory (Chao and Gross, 1987; Chao et al., 1995; Chao and Ding, 2016; Xu and Chao, 2017). One can then calculate the earthquake-induced $\Delta E_g$ by equating it to the work done by $\mathbf{u}(\mathbf{r})$ against the local gravitational acceleration $\mathbf{g}(\mathbf{r})$ integrated over the volume of the self-gravitating Earth:

$$\Delta E_g = -\iiint \rho(\mathbf{r})\mathbf{u}(\mathbf{r}) \cdot \mathbf{g}(\mathbf{r})\, dV, \tag{1}$$

where $\rho$ is the mass density. In the approximate spherically symmetric monopolar Earth, the proper formula for $\mathbf{g}(\mathbf{r})$ turns Equation (1) into:

$$\Delta E_g = 16\pi^2 G \int_0^R \rho(r)\, u_r(r) \left[\int_0^r r'^2 \rho(r') dr'\right] dr, \qquad (2)$$

where only $u_r$ (positive outward), the radial component of $\mathbf{u}$, makes contribution. We point out that Equation (2) corresponds to the monopolar term of equation (12) of Chao et al. (2026) describing the layer-by-spherical-layer mass accretion from infinity. The ultimate fate of the mass-accretion $E_g$ is converting into heat; in fact, mass accretion represents the process of maximum $E_g$ release subject to the conservation of momentum.

Using the GCMT Catalogue containing all major earthquake events greater than Mw ~5.0 since 1977 (numbering over 45,000 by 2016), what have been asserted quantitatively (Chao, 1995; Chao, and Ding, 2016) include: (i) The earthquake-induced $\Delta E_g$ is typically ~1000 times greater than the seismic energy, an assertion first conveyed by Dahlen (1977). In an extreme example the 2004 Sumatra event singly released ~3 × $10^{21}$ J of $E_g$, or ~2.1 years' worth of the global terrestrial heat flow of ~45 TW (terawatt). (ii) Thrust-faulting events (with positive isotropic component $M_{rr}$ of the source seismic moment tensor) give negative $\Delta E_g$, meaning they induce releases of the Earth's $E_g$. Contrariwise, normal-faulting events (with negative $M_{rr}$) give positive $\Delta E_g$, meaning they increase $E_g$. (iii) On Earth the thrust-faulting events induce an average $E_g$ release of power ~10 TW, nearly a quarter of the total terrestrial heat flow and far overshowing in prevalence and seismic magnitude the normal-faulting events with average induced $E_g$ gain of power ~2 TW.

Meanwhile it is well known that the thrust-faulting earthquakes occur primarily along the world's subduction zones at converging plate boundaries, whilst the normal-faultings along the spreading centers at divergent boundaries. The inference of the net $E_g$ balance of the two, –10 + 2 = –8 TW, being negative is consistent with the spontaneous thermodynamic evolution of a physical system seeking lowest energy state as a whole. It is noted that the said $E_g$ does not belong to the earthquake; it belongs to the gravitating Earth itself. The earthquakes merely induce changes in it.

We now focus on the dominant happening in the above scenario, namely the release of as much as 10 TW of $E_g$ at the subduction zones. What becomes of this released $E_g$ at depth? The most natural fate in keeping with the second law of thermodynamics of increasing entropy is in the form of in-situ heat. Take for example a piece of rock held mid-air by some mechanism in force-balance which suddenly fails, and the rock falls spontaneously until hitting the floor where it rests back to another force-balance state. The mechanical failure represents the earthquake faulting; during the fall the rock creates air friction and turbulences and emits sound or "seismic" waves. But the dominant happening energy-wise is the eventual and total conversion of the original excess $E_g$ into heat irreversibly upon the rock's impact with the floor, symbolizing a mass accretion process.

Dahlen (1977), however, by enforcing the hydrostatic equilibrium condition, took the before-and-after force-balance states to be the same state, somehow reversibly maintained throughout the faulting process. It was thereby argued that any change in gravitational energy, of however great amount, is inconsequential because it essentially gets exchanged back to the form of stored elastic energy, essentially entropy-free in totality awaiting next faulting. In contrast, as a key point here, we maintain that a real earthquake is the extreme opposite to Dahlen's "idealized earthquake" scenario. First of all, no earthquake faulting would even happen under a hydrostatic equilibrium condition in the first place. The very fact that it does happen signifies the start of a complete failure of the equilibrium condition, whereby the inertial term is not negligible and the equilibrium equation fails. Like the fallen rock, the irreversible process lasts until reinstated into a new hydrostatic equilibrium state of a lower $E_g$ awaiting next faulting. The new state differing from the initial state is clearly and amply evidenced by observations from GNSS and InSAR (for surface geometric deformations) and GRACE satellites (for time-variable gravity): The earthquake-induced displacement field $\mathbf{u}(t)$ is a jump (as in a step function) as opposed to an impulse (as in a delta function) in time. Nothing reverts reversibly back to the initial energetic state co-seismically (or post-seismically for that matter), as surmised in Dahlen's equilibrium scenario. More importantly in the present context, the latter equilibrium scenario is inconsistent with the evolving nature of the thermodynamic Earth in increasing entropy, and contradicting the energetic role played by the earthquakes in plate tectonics.

We now address the ensuing questions: What becomes subsequently of the regional heat now converted from the earthquake-induced $\Delta E_g$? And how does this excess $E_g$ originate in the first place?

## 3. Gravitational Energy in Mantle Convection

First we examine the energetic cycle in a generic 2-D Rayleigh-Benard convection of a thermally compressible fluid under gravity, in a rectangular walled box between two heat reservoirs: the hot bottom and the cold top. We assume steady state with motions slow enough that the inertia term can be neglected but fast enough to obey adiabaticity in the vertical limbs of descending (by positive gravity) and ascending (by buoyancy, or negative gravity). In both limbs, the $E_g$-heat conversion is governed by the relation that the viscous dissipation equals the release of $E_g$ (e.g., Morgan et al., 2016). The viscous dissipation depends on the fluid viscosity times the relative shear velocity, ending up as heat through molecular friction. The $E_g$ release is proportional to the density change (on two ends of the limb) times the vertical velocity (up is positive) (Jaupart et al., 2015; Morgan et al., 2016). The density change, in turn, is proportional to the (positive) thermal expansion times the super-adiabatic temperature departure of the fluid parcel. Hence, in both limbs where the critical Rayleigh number had been exceeded, the spontaneous $E_g$ change has the same sign -- being released to viscous dissipation either descending or ascending. The descending is likened to a piece of rock falling in a viscous liquid (say honey), whereas the ascending is likened to a piece of Styrofoam rising, or “falling upward”, when released from the bottom of a tank of honey. In the latter case, just as in the former, the system loses $E_g$ because of the buoyancy force as negative gravity, which now does mechanical work while the fluid parcel is falling upward.

The two anti-symmetric limbs of course do not comprise a perpetual machine, because the two horizontal branches of the convection cell connecting them carry out heat exchanges with the basal hot and the top cold reservoirs to operate the thermodynamic machine. That resets the original state for both “falling” processes. In an alternative perspective, it can be construed that the two horizontal branches replenish $E_g$ (Morgan et al., 2016), awaiting the upcoming “fall”. The bottom branch replenishes the stored $E_g$ by the mechanical work done through thermal expansion against the ambient pressure, whereas the top branch does so by thermal contracting receiving work done by the ambience. That constitutes an anti-symmetric thermal process counterpart to the vertical limbs in the convection cycle. Of course such (adiabatic) expansion/contraction involves work done in the vertical limbs as well, being an implicit budget item in the overall $E_g$-heat conversion.

Thanks to its configuration, the mantle convection departs from an ideal Rayleigh-Benard convection in significant ways (Turcotte and Shubert, 2002; Anderson, 2007; Davies, 2013; Jaupart et al., 2015; King, 2015; van Keken and Wilson, 2023; Davies et al., 1992; Stacey and Davies, 1992). The mantle itself generates considerable extra amount of heat from the radiogenic sources. The radiogenic heat, of uncertain amount and spatial distribution, is dissipated throughout the mantle including ascending and descending regions (in this sense spoiling the ideal adiabaticity). That reduces the strength of the ascending which now does so in a heated environment, thus decreasing their thermal buoyancy (Deschamps et al., 2010). As a result, convection gets progressively dominated by descending currents with increased amount of radiogenic heating. Physical properties of mantle rocks may alter the distribution of viscous dissipation between descending and ascending limbs. In particular, the strong dependence of viscosity on temperature implies that slabs are much more viscous than plumes. In terms of forcing, the plate tectonic process is consistent with a top-down mantle convection driven by gravitation, thereby the gravitational slab pull at subduction overtakes the ridge push at the spreading end by a factor over ~2 thanks to its enhanced adiabatic temperature difference. The existence of the upper-lower mantle transition at 660 km depth due to material phase changes acts as an obstruction to the subducting slab and further decreasing the buoyancy of plumes, bearing on the whole-mantle versus multi-layered convection cells. Together with the viscosity increase from lower to upper mantle, this may result in (temporary) slab stacking around 660-1000 km, and to plume ponding below the 660 km transition. Added to the complication are the Earth’s spherical (as opposed to rectangular) geometry, the no-vertical-wall boundary conditions, the radially decreasing $|\mathbf{g}|$ (albeit not by much) through the mantle, and the presence of the hot-spot mantle plums emerging from the deep whether actively or passively.

These deviations from an ideal Rayleigh-Benard convection result in the observation that the subduction is by-far the most vigorous segment of the mantle convection, on which we focus next along with the solidity of the subducting slab.

## 4. The Island-Arc Volcanism Paradox

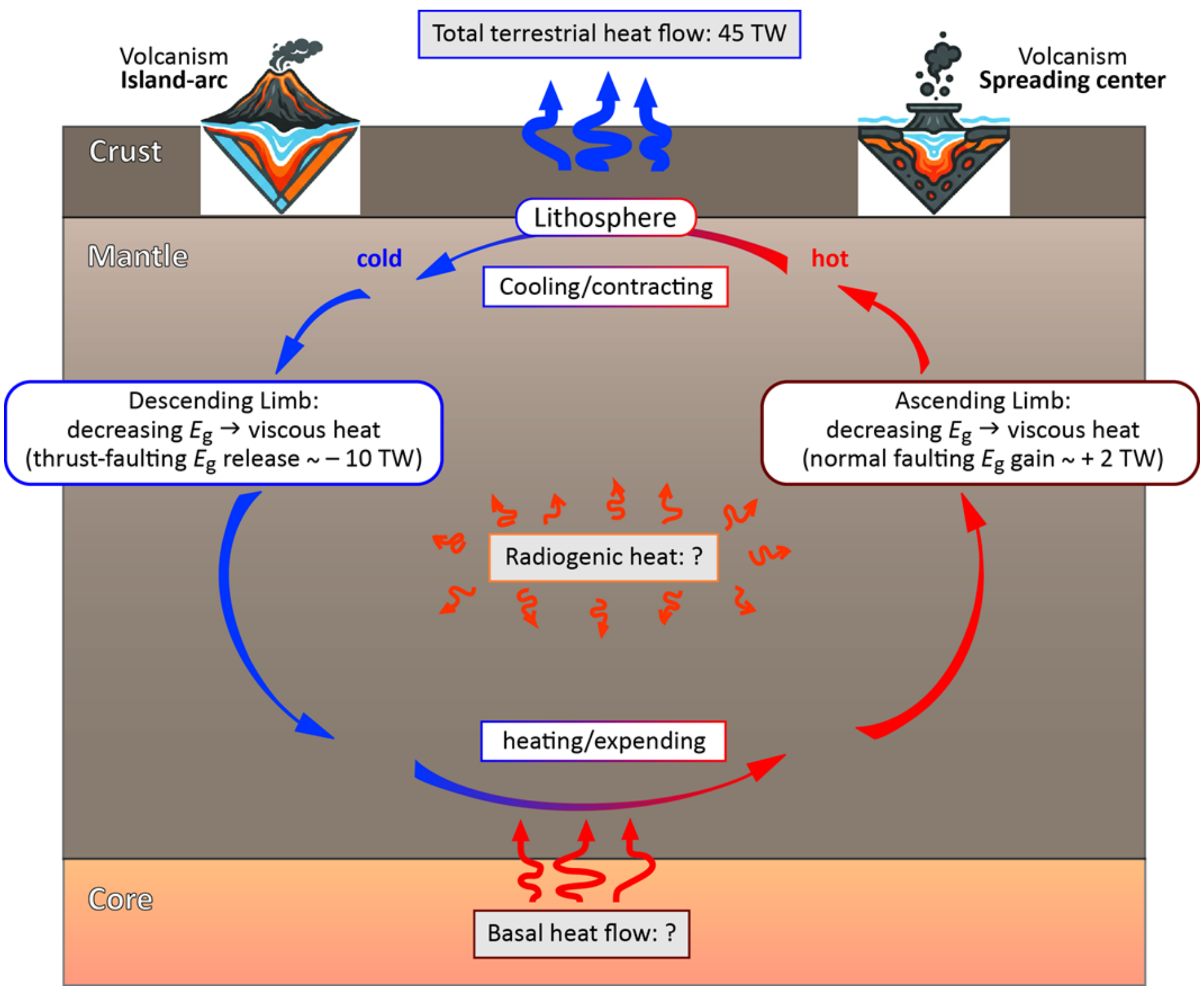

Fig. 1. The major $E_g$-heat exchange cycle within a generic mantle convection cell, depicting the hot ascent (as red) below the divergent spreading center versus the cool descent (as blue) of the convergent subduction zone, along with respective volcanisms. The subduction zone (upper-left corner of the cell) constitutes where thrust-faulting earthquakes play the major role of converting $E_g$ into heat in the lithosphere to supply for the island-arc volcanism, at power ~10 TW, nearly a quarter of the total terrestrial heat flow of ~45 TW.

Figure 1 depicts a generic mantle convection cell while incorporating the earthquake-induced $\Delta E_g$ according to Equation (1) and the major energetics of the $E_g$-heat exchange during the irreversible thermodynamic cycle.

Figure 2(a) displays a compilation of historical global heat flow measurements (adopted from Davies, 2013). The unit heat flux in oceanic areas being generally higher than on continents by a factor of ~2. Figure 2(b) (courtesy of USGS and US National Park Service) shows the global locations of known volcanoes. The subduction zones along the mighty Pacific Ring of Fire (including the Indonesia island chain) and western Mediterranean see: (i) unequivocal high heat-flow anomalies; (ii) intense island-arc volcanoes; (iii) strong thrust-faulting earthquakes as afore-mentioned (while the spreading centers see the weaker normal-faulting seismicity).

The accompanying seismicity and volcanism at the subduction represent by-products that are construed independent mechanically: The earthquakes as brittle breaks occur to accommodate the subducting plate movements, regardless of the volcanism. Reciprocally, the volcanism occurs thermodynamically on itself, in principle independent of seismicity.

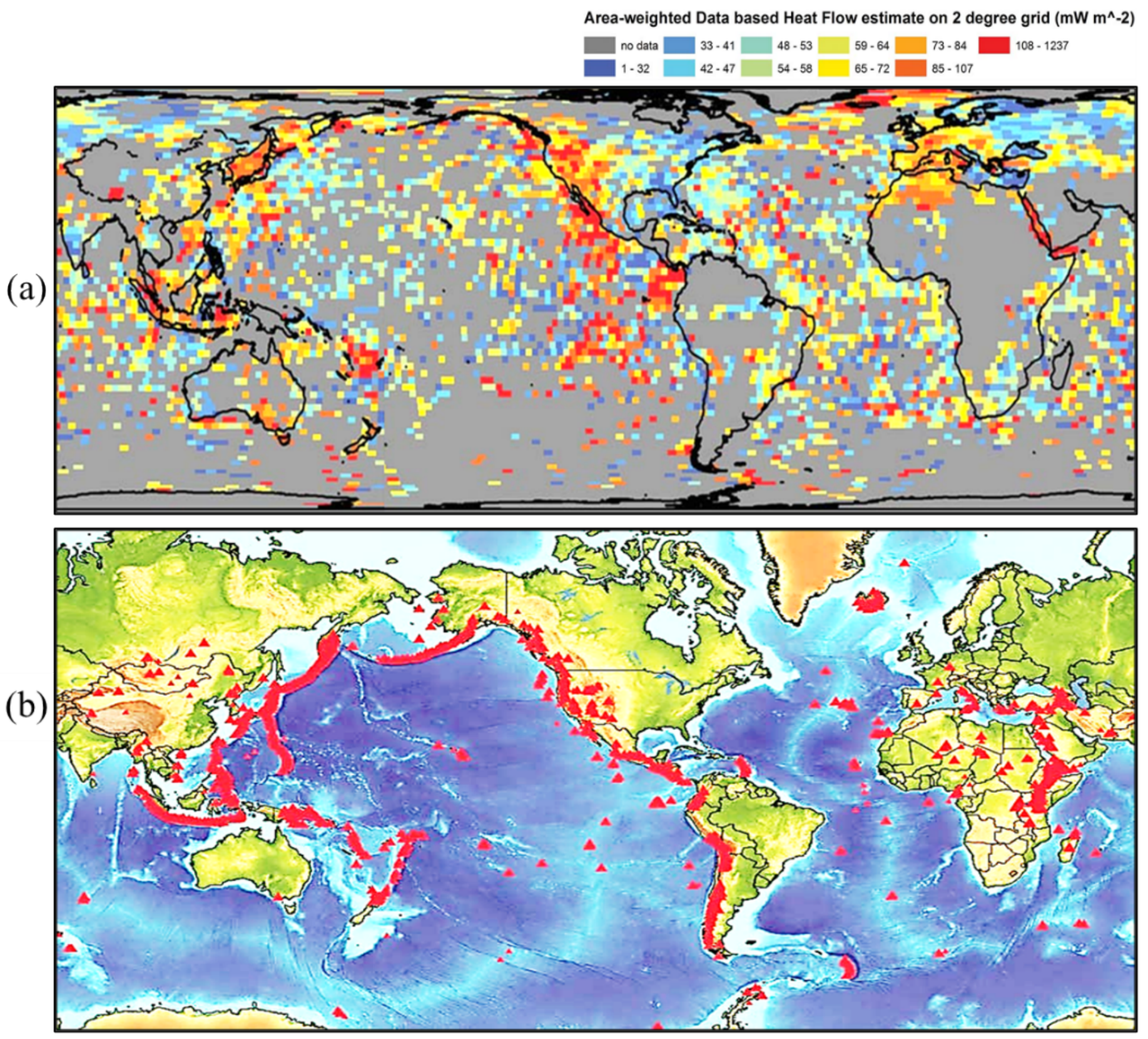

Fig. 2. (a) Global distribution of the terrestrial heat flow on a 2° equal-area grid (duplicated and recast from Davies, 2013), showing the high heat flow anomalies (reddish areas) at the subduction zones along the Pacific Ring of Fire (including the Indonesia island chain) and western Mediterranean, as well as at the spreading centers. (b) Geographical locations of world's known volcanoes (red triangles; courtesy of USGS and US National Park Service) against their plate tectonic setting, highlighting majorly the island-arc volcanism coinciding with the seismic belts (not shown) associated with the subduction.

Here arises a long-standing paradox in plate tectonics (McKenzie, 1969; Stern, 2002; Turcotte and Shubert, 2002; Anderson, 2007; King, 2015): What supplies the heat for the intense island-arc volcanism at the subduction zones, the supposedly cool end of the mantle convection? The island-arc volcanism does necessitate melting of subducted material as magma. For facilitating the melting process in lack of ready sources for heat, various physical and chemical mechanisms have been proposed in the literature (e.g., Davies and Stevenson, 1992; Stern, 2002; Jaupart e al., 2015; van Keken and Wilson, 2023), including the melting point depression due to the presence of water in the subducted material, shear friction heating via viscosity, mantle wedge dynamic flows, adiabatic and latent heat, material phase transitions, etc. While able to facilitate the magma generation and heat transports that help configure the island-arc volcanoes to the way they are, these proposed passive processes do not address the *excess* heat nor its source(s). The melting itself does not mean extra heat, and conversely the physical heat source has little to do with whether or not the magma is melted. An analogy can be drawn in the de-icing practice where salts are applied on roads to lower ice's melting/freezing point, while the process adds no net heat to the system and involves little temperature change, if at all. At any rate the volcanism is not a reason, but a consequence, of the excess heat.

Tying together all the above, we surmise the following scenario that stands to resolve the island-arc volcanism paradox:

In the descending limb of subduction zone (upper left corner of Fig. 1), the subducting slab needs to dissipate the (insensible) $E_g$ in the lithosphere into (sensible) heat, which Morgan et al. (2016) estimated to be ~11-14 TW. That, however, can hardly happen on itself because this cool end of convection carrying along the isotherms down-dip is inconducive to viscous dissipation. Instead of viscous flows, the cool, brittle subducting lithosphere undergoes breaks as thrust-faulting earthquakes, which constitute the primary physical mechanism to accomplish the said $E_g$-heat conversion, along with any incumbent adiabatic, latent and conductive heat exchanges. Upon conversion into heat at the power of 10 TW, the earthquake-induced $E_g$ release supplies amply for the island-arc volcanism, as a result manifesting as the high heat-flow anomalies at subduction zones shown in Fig. 2(a).

Meanwhile in the ascending limb or spreading center (upper right corner of Fig. 1) the bulk of $E_g$-heat conversion at ~half of power as in the descending limb (Morgan et al., 2016) also suffices to supply for the spreading-center volcanism (shown in Fig, 2a) as the ascending material undergoes decompression and thinning when approaching surface. The rest (equivalent to the "efficiency" of a heat engine) remains or reverts back to $E_g$, including the portion reclaimed via the normal-faultings per Equation (1) at the lower power ~2 TW (positive $\Delta E_g$) as afore-mentioned, which is incorporated into the $E_g$ storage in the cooling lithosphere to be re-released later when descending.

## 5. Discussion and Conclusions

### 5.1 Secondary effects

Certain secondary factors in the convection-seismicity-volcanism energetic connection have been ignored. Besides the thrust- and normal-faultings, the strike-slip earthquakes being largely horizontal and involving far lesser $\Delta E_g$ are rarer and smaller anyway. Similarly ignored are the seismicity associated with back-arc spreading (hence not thrust-faulting) at subductions and transform faults (hence strike-slip) at spreading centers. Similarly the energy involved in the hot-spot plume volcanism is at least one order of magnitude smaller than those of spreading centers and subduction zones, a subject perhaps for elsewhere.

The global thrust-faulting earthquakes (overtaking the normal-faultings) have been found to steadily make the Earth rounder (less oblate) and more compact, all indicative of the said long-term thermodynamic evolution. They induce a secular *increase* of Earth's rotational, or more precisely the spin, kinetic energy when making the Earth less oblate and spin faster under the conservation of angular momentum (Dahlen, 1977; Chao and Gross, 1995; Chao et al., 2026). Under the equivalence principle, that amounts only to the power typically ~1/300 that of $E_g$ (thus about 3 times larger than the seismic energy), because the spin centrifugal acceleration for $\mathbf{u}$ to work against is ~1/300 of the gravitational acceleration on Earth's surface (cf. Equation 1).

### 5.2 Spatial extent of earthquake-induced $\Delta E_g$

The relatively concentrated heat-flow anomalies around the subduction zones in Fig. 2(a) implies near-surface energetic processes. We inspect the depth extent of the earthquake-induced $\Delta E_g$, where the $\Delta E_g$-heat conversion occurs. The radial integration kernel (to be multiplied with $M_{rr}$ which is positive for thrust-faulting and negative for normal-faulting) for evaluating the co-seismic $\Delta E_g$ (by Equation 1) takes positive values at depth above the earthquake focus, whereas negative values below the focus and through the upper mantle while tapering off into deeper mantle (Tanimoto and Okamoto, 2000; Chao and Ding, 2016; Xu and Chao, 2017). The latter dominates the former typically by an order of magnitude yielding the net effect of sorts discussed above. Thus the overall earthquake-induced $\Delta E_g$ release at the subduction zones happens predominantly in the upper mantle within several hundred km depth-wise. This also suggests that the *lateral* extent of the $\Delta E_g$-heat conversion happens within the width on the order of a thousand km along the island arcs, which gets further dispersed spatially with time.

The said depth entrainment of the integration kernel for $\Delta E_g$ into the mantle owes to the elasticity of the mantle co-seismically. It follows that the *anelastic* post-seismic and inter-seismic surficial displacements, the notion of apparent earthquake cycles, stay merely surficial and hence deprived of much gravitational effect depth-wise in comparison to the co-seismic (Xu and Chao, 2019). If anything, instead of cycling back, they would *augment* to the co-seismic effect as part of the long-term irreversible process of the mantle convection engine. The same argument incidentally applies to other low-degree geodynamic quantities including the Earth's volume, moments of inertia, and rotation (Chao and Ding, 2016).

The end effect is consistent with the observation that the upper mantle sees far more active convection than the lower mantle. We have intentionally shunned specifics regarding the bottom portion of the convection, whether it be the core-mantle boundary or the 660 km upper-lower mantle transition level below which the subduction process may encounter temporary slab stacking or material ponding. Particular deep-focus earthquakes release $\Delta E_g$ reaching depths well in the lower mantle, nevertheless the ensuing heat transfer carries on in long term despite that. In fact, seismicity being a surficial dynamic process, our scenario is not particularly sensitive to whether the convection is whole-mantle or multi-layered or hybrid. Conversely, our scenario offer limited constraints on the deeper convection pattern involving, for instance, the D" layer which is thought to be the signature of the phase transition from bridgmanite to post-perovskite, or the LLVPs (large low-velocity provinces) which are likely hotter and chemically differentiated compared to the surrounding mantle.

### 5.3 New paradigm for the role of earthquakes in plate tectonics

The 21$^{st}$ century mega-seismicity is evidently undergoing an activity cycle since the previous intense episode of 1950-60s. We suppose that our earthquake-induced $\Delta E_g$ power estimates are representative of the present-era mean. Notably, they are calculated based on sound seismological theory (normal mode or elastic dislocation theories) along with well-proven knowledge about the Earth, yet independent of any thermodynamic modeling. It is remarkable, and physically significant then, that the earthquake-induced $\Delta E_g$ powers are found to closely match those made per thermodynamic arguments for the mantle convection (Morgan et al., 2016), both in quantity and in physical characteristics as summarized in Fig. 1.

Dictated by plate tectonics the spatial distribution and source mechanisms of global earthquakes are fairly well understood in terms of the system forcing. In parallel, the *energetic* processes involved reflect the same ultimate physics but in a distinct light, putting forth the following new paradigm as to the identity and dynamic role of earthquakes in the grand scheme of plate tectonics:

At the subduction zones, earthquake faultings are bestowed the responsibility of releasing the stored $E_g$ during slab descents, which they indeed duly accomplish with the thrust type of focal mechanism. While an earthquake's seismic magnitude is defined per the elastic stress drop, energetically the collective sizes of these earthquakes must amount to ~10 TW of $E_g$ release rate as necessitated by the mantle convection's descending process. In that sense, $\Delta E_g$ proves to be a deciding factor for not only the focal mechanisms but also the sizes of the thrust-faulting earthquakes. The $E_g$ release both supplies for and lend constraints on the overall sizes entitled by the earthquakes, to the extent that their long-term mean magnitudes can be anticipated. Conversely the excess heat at subduction zones owes its existence directly to earthquake activities, in such a way that the present-day heat flow represents a time-lapse record of $\Delta E_g$ release induced by long-past seismicity. The same arguments apply to the ascending process at spreading centers as well, where opposite and less vigorous normal-faulting earthquakes build up $E_g$ at ~2 TW.

Thus, in a broadened perspective the global earthquakes are not just passive consequences, but active, integral parts of the plate tectonics in keeping with the mantle $E_g$ energetics and its evolution. A conjecture then arises whether the global seismicity has actually been an active *facilitator* that quickens the $E_g$-heat conversion more efficiently than otherwise. As such, the earthquakes participate in a positive feedback (Chao et al., 1995) by facilitating the mantle convection and hence plate motions which in turn give rise to seismicity. The opposite is equally conceivable as far as forcing is concerned, that seismicity *impedes* the $E_g$-heat exchanges via fault locking, forming an energetic negative feedback. Either way the global seismicity influences the ultimate rate of internal energy transports and thus the mechanisms of the long-term cooling of the Earth. This plate tectonics-based perspective naturally pertains to the period of the tectonic mode in the Earth's cooling history (26). It'd be interesting to learn whether these mechanisms are unique to the Earth or reminiscent to other terrestrial planets and moons.

**5.4 Self-consistency in numerical modelling of slab subduction**

Finally, it can be suggested that when established the above physical scenario be incorporated into the numerical modelling of mantle convection and particularly the subduction processes (cf. van Zelst et al., 2019). As is typical, such models (Davies and Stevenson, 1992; Syracuse et al., 2010; King, 2015; Zhong et al., 2015), kinematic or dynamic, pay far less attention to energy than to forcing. When energy is explicitly treated, the energy-relevant quantities are often prescribed a priori in terms of the temperature field with the radiogenic and latent/chemical energy release as the obvious heat source candidates. Here we maintain that, while typically overlooked in the numerical models, $E_g$ ought to be tracked thermodynamically through the convection cycles, as the viscous and frictional heat typically incorporated in the models in the presence of the velocity field.is a product of the $E_g$-heat conversion. It presumably also bears on the quest (Korenaga, 2003) of the supposedly extra heat reservoir(s) yet unaccounted for in mantle energetics. At any rate, it should be proper to incorporate into the thermodynamic modeling the earthquake-induced $\Delta E_g$, considering its significance and active participation in the subduction process and volcanism.

Acknowledgments. This work is supported by the National Natural Science Foundation of China (grant #42530114), and additionally by the National Science and Technology Council (grants #113-2116-M-001-019 and #113-2116-M-001-021) and Academia Sinica Investigator Project (grant #AS-IA-113-M02) of Taiwan. We thank C. Xu and W. Kuang for insightful discussions. The earthquake-induced gravitational energy changes cited in this work were previously published based on the global Centroid Moment Tensor (CMT) catalogue (http://www.globalcmt.org/CMTsearch.html). Fig. 2(a) is duplicated and recast from Davies (2013); Figure 2 (b) is courtesy of USGS and US National Park Service (https://www.nps.gov/articles/volcano-monitoring.htm).